\documentclass[aps,prl,twocolumn,showpacs,floatfix,superscriptaddress]{revtex4}
\usepackage[]{graphicx}
\usepackage{epstopdf}
\usepackage{graphicx}% Include figure files
\usepackage{dcolumn}% Align table columns on decimal point
\usepackage{bm}% bold math
\usepackage[usenames]{color}    % comments and corrections in colors

\usepackage{siunitx}
\usepackage{enumerate}

\begin{document}

\title{Phase retrieval enhanced by quantum correlation}

\author{Giuseppe Ortolano}
\email{giuseppe.ortolano@polito.it}
\affiliation{Quantum metrology and nano technologies division,  INRiM,  Strada delle Cacce 91, 10153 Torino, Italy}
\affiliation{DISAT, Politecnico di Torino, Corso Duca degli Abruzzi 24,
10129 Torino, Italy}

\author{Pauline Boucher}
\author{Ivano Ruo Berchera}
\affiliation{Quantum metrology and nano technologies division,  INRiM,  Strada delle Cacce 91, 10153 Torino, Italy}

\author{Silvania F. Pereira}
\affiliation{ Imaging Physics Dept. Optics Research Group, Faculty of Applied Sciences, Delft University of Technology, Lorentzweg 1, 2628CJ Delft, The Netherlands}

\author{Marco Genovese}
\affiliation{Quantum metrology and nano technologies division,  INRiM,  Strada delle Cacce 91, 10153 Torino, Italy}
\begin{abstract}
Quantum correlation, such as entanglement and squeezing have shown to improve phase estimation in interferometric setups on one side,  and non-interferometric imaging scheme of amplitude object on the other. In the last case, quantum correlation among a pair of beams leads to a sub-shot-noise readout of the image intensity pattern, where weak details, otherwise hidden in the noise, can be appreciated. In this paper we propose a technique which exploits entanglement to enhance quantitative phase retrieval of an object in a non-interferometric setting, i.e only measuring the propagated intensity pattern after interaction with the object. The method exploits existing technology, it operates in wide field mode, so does not require time consuming raster scanning and can operate with small spatial coherence of the incident field. This protocol can find application in optical microscopy and X-ray imaging, reducing the photon dose necessary to achieve a fixed signal-to-noise ratio.	
\end{abstract}	

\maketitle

\section*{Introduction} 

Quantum imaging \cite{Berchera_2019} and sensing \cite{Pirandola_2018,Degen_2017} are important and relatively developed sub-field of quantum technologies that can lead to a new generation of measurement instruments, with applications ranging from fundamental physics \cite{Aasi_2013,Pradyumna_2020}, to bioimaging \cite{Taylor_2016} and microscopy \cite{Samantaray_2017,Casacio_2021,Monticone_2014}. At the base of those applications there are the fundamental problems of phase and loss estimation making use of quantum resources, that have been extensively studied showing the possibility to beat classical limit in the accuracy. Phase estimation in interferometry exploiting squeezing and entanglement can lead to approach in principle the Heisenberg limit \cite{Giovannetti_2011,Polino_2020}, i.e. a reduction of the estimation uncertainty of $N^{-\frac{1}{2}}$ with respect to the standard quantum limit, where $N$ is the number of probing particles. However, in realistic scenario, such as in presence of detection losses and decoherence, the improvement is rather in terms of by a constant factor, i.e. the scaling with the number of particles is not affected \cite{Demkowicz_2014}. Concerning the estimation of a loss parameter $0\leq\tau\leq1$, such as the transmittance or reflection from a sample, the ultimate quantum limit attainable with quantum probe is given by $U_{uql}=[\tau(1-\tau) /N]^{\frac{1}{2}}$  \cite{Jakeman_1986,Monras_2007,Adesso_2009,Nair_2018}, providing an advantage of $\sqrt{\tau}$ over the uncertainty of the best possible schemes limited to classical states. Note that, for low transmittance of the system, as it is for thin biological or nanofilms, this advantage can be disruptive. Both quantum phase and loss estimation have been turned to imaging scheme, where the multiparameter 2-D spatial profile of a system has been reconstructed \cite{Ono_2013, Israel_2014, Frascella_2019}. Among them, some imaging schemes are of particular interest since they can work in wide field mode, meaning that the image of the object is obtained in one shot, without raster scanning. In sub-shot-noise quantum imaging SSNQI reported in \cite{Brida_2010,Samantaray_2017} a weak absorbing object is probed by one of the two beams produced by spontaneous parametric downconversion (SPDC) and imaged at a pixel array of a high quantum efficiency camera. The second beam is spatially quantum correlated point-by point in the photon number fluctuation and it is detected separately, for example in another area of the camera. In this way, the same noise pattern that affects the image is measured on the second beam and by properly subtracting the two beams from one another, the ideal (unit detection efficiency $\eta=1$) reconstruction of the 2-D transmission profile of the object can be achieved, reaching the $U_{uql}$ \cite{Losero_2018,Berchera_2020}.

In this paper we will show that a  a scheme similar to the SSNQI for amplitude object can be used also to extract phase information in a more efficient way compared to the classical case thanks to the quantum correlations: we will name this protocol quantum correlation (enhanced) phase retrieval (QCPS). Its working principle relies on the phase induced propagation effect on the intensity pattern, so it is not an interferometric estimation. The seminal idea has been suggested by some of the authors of this article \cite{Ortolano_2019} for imaging of refractive objects, but the model was over-simplified without considering specific (and realistic) phase induced propagation effects.  Here, we consider a feasible reconstruction algorithm based on the solution of the so called transport of intensity equation (TIE) which leads to a unique and quantitative wide field reconstruction of a phase profile \cite{Teague_1983,Paganin_1998,Zuo_2020}. However, the reconstruction can be strongly affected by the noise, thus we investigate how quantum noise reduction in our scheme boots the quality of the retrieval. QCPS can work with partially coherent light and has some advantage with respect to interferometric scheme: it can be directly applied to wide field transmission microscopy and it is intrinsically more stable than an interferometric setup \cite{Zuo_2020}.

\section{Materials and Methods}

 \subsection*{Phase retrieval by TIE.} 
 	A non interferometric method \cite{Teague_1983} to retrieve the phase of an object consists in the measurement of the intensity $I(\bm{x},z=0)$ at the object plane of coordinate $\bm{x}$ and its derivative along the propagation axis $z$. The derivative is computed by a finite difference of two measurement out of focus of a distance $\delta z$, $\frac{\partial}{\partial z}I(\bm{x},z)\approx\Delta I(\bm{x},\delta z)/(2 \delta z)$ with  $\Delta I(\bm{x},\delta z)= I(\bm{x},\delta z)-I(\bm{x},-\delta z)$. Under paraxial approximation, the phase is then retrieved by means of the TIE:
 \begin{equation}
 -k\frac{\partial}{\partial z}I(\bm{x},z)=\nabla_{\bm{x}}\cdot\left[I(\bm{x},0)\nabla \phi(\bm{x},0)\right]
 \label{tie}
 \end{equation} 
 
 Using energy conservation considerations, this equation has been proven valid even with partially coherent sources \cite{Paganin_1998}. This feature makes the TIE approach perfectly suited for being used with light from SPDC, where transverse and longitudinal coherence length can be much smaller than the object size and the whole illuminating beam. This is not a secondary aspect, since it is exactly due to the multimode nature of the emission that correlation shows a local character and shot noise can be removed pixel-by- pixel in the image. The solution of the Eq. (\ref{tie}) is unique provided that the on focus intensity $I(\bm{x},0)$ and the intensity derivative along $z$ are known and the phase is continuous.\\ 
 Since in the scheme the derivative along $z$ is estimated by a finite difference, smaller defocus $\delta z$ would lead to a more precise approximation of the local derivative and in turn to a better estimation of the phase. However, smaller values of $\delta z$ lead to increasingly smaller values on the measured finite difference $\Delta I$, with a low signal to noise ratio in the presence of unavoidable noise, technical noise but also of fundamental origin such as the shot noise.  On the other side, increasing the defocus distance can reduce the effect of noise but it can be done up to a point because the linear approximation of the derivative in Eq. (\ref{tie}) may not be longer valid, and the first effect can be a loss of the higher frequency component of the phase profile. In general this two competing trends lead to an optimal value of the defocus distance $dz$ which depends on the noise level and on the phase function itself. Those aspects, in particular the problem of noise in the phase retrieval by TIE are extensively and quantitatively analyzed in \cite{Paganin_2004}. First, we assume that the intensity is varying sufficiently slowly that the intensity derivative is dominated by the effects of phase curvature, so that the right side of Eq. (\ref{tie}) can be safely approximated as $I_{0}\nabla^2\phi(\bm{x},0)$. Then, we consider for a moment that the only contribution to the finite difference $\Delta I(\bm{x},\delta z)$ is the noise fluctuation on the intensity measurement, $\sigma(\bm{x})$ . In this case, substituting the latter in Eq. (\ref{tie}), one has that the phase artifacts in the reconstruction due to the noise are:
  \begin{equation}
 -k\frac{\sigma(\bm{x})}{\sqrt{2} I_{0} \delta z}= \nabla_{\bm{x}}^2\phi_{noise}(\bm{x}).
 \label{noise artifact 1}
 \end{equation}

 The noise is assumed independent in the two plans $+\delta z$ and $- \delta z$, so it has been combined in quadrature. The Eq. (\ref{noise artifact 1}) can be solved by taking the Fourier transform on both sides, leading to
 
   \begin{equation}
 k\frac{\tilde{\sigma}(\bm{q})}{4 \pi^{2}\sqrt{2} I_{0} \delta z |\bm{q}|^{2}}= \tilde{\phi}_{noise}(\bm{q})
 \label{noise artifact 2}
 \end{equation} 
  where te tilde indicate the Fourier transform and $\bm{q}$ is the spatial frequency. The damping factor $|\bm{q}|^{2}$ of the higher frequencies  at the denominator of Eq.   (\ref{noise artifact 2}) and the fact that the quantum noise (shot noise) has a flat white spectrum $\sigma_{SN} (\bm{q})= \sigma_{SN}$, indicate that the effect of shot noise is to generate artifact especially at lower frequency which are not intrinsically suppressed by the phase retrieval algorithm. Those low frequency noise will be evident in the simulations presented in the 'Results' section. Moreover, in the direct problem of propagation, higher frequency of the phase object generate stronger effect on the intensity. Thus, based on these remarks, the regions with rapid changes in the phase (higher frequency) are better reconstructed than the ones characterized by slow curvature.

\subsection*{QCPS: the scheme}

	\begin{figure}[h]
		\centering
		\includegraphics[width=0.5\textwidth]{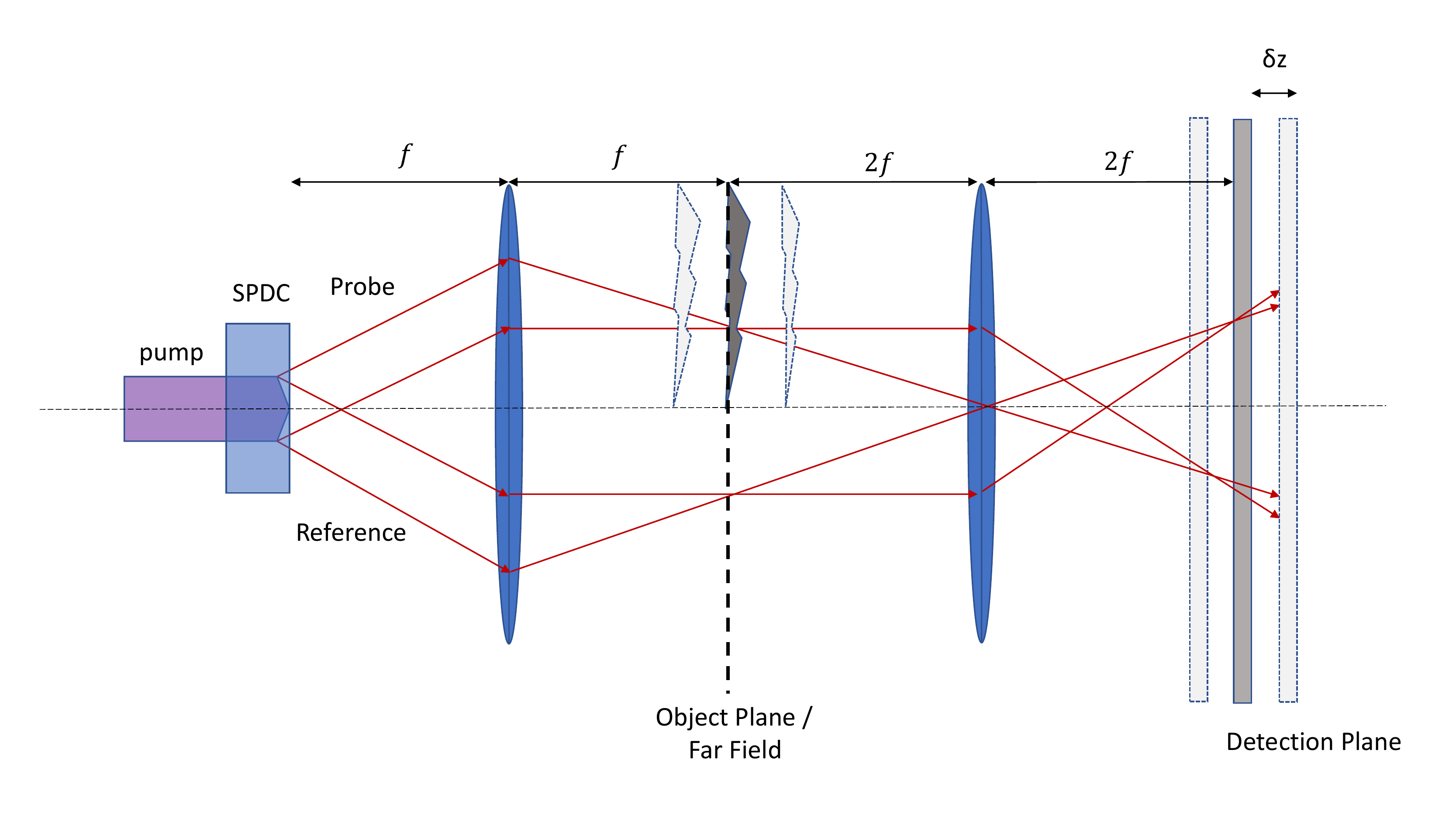} 
		\caption{\emph{QCPR}. Two beams, labeled probe and reference, produced by an SPDC source, propagate through an imaging system composed by two lenses and a test object. The object is placed in the far field of the source, and only the probe beam interacts with it. A suitable imaging system ($2f-2f$ in the picture) projects the image of the object plane at the camera chip. Phase information can be retrieved from intensity measurements out of focus. The defocus can be achieved either by a shift of the detection plane or a shift of the object.}\label{scheme}
	\end{figure}

	The proposed QCPR protocol exploits the scheme depicted in Fig. \ref{scheme}. We consider a source of two quantum correlated beams, namely the SPDC process, producing two intensity patterns that are perfectly identical in the far-field, where the pure phase object is placed. Even the shot noise component of the noise is perfectly reproduced in the two beams, something that is not possible in the classical domain. One of the beam probes the object, while the other is used as the reference for the noise. In fact, the acquisition of the noise pattern measured on the reference beam allows to remove it from the image of the probe perturbed by the object. The same noise reduction is applied for each of the three intensity patterns needed for the phase retrieval algorithm, i.e. for the plane at $z=0, +\delta z, -\delta z$. In this way, noise-induced artifacts of the reconstructed phase are expected to be strongly reduced. 
	
	In the following we will investigate the details of the scheme, first describing the correlation properties of the source and than modeling an efficient way to use them for the noise reduction.
\subsection*{SPDC photon statistics and scale dependent correlations}	
	 SPDC in the low gain regime is usually described as a process in which the photons of a pump beam (p), thanks to the interaction with a non-linear crystal, have a small probability to convert in a couple of photons, usually called signal (s) end idler (i), subject to conservation of energy, $\omega_p =  \omega_s + \omega_i$, and the momentum, $\textbf{k}_p =  \textbf{k}_s + \textbf{k}_i$. Thus, under the plane wave pump approximation signal and idler photons are perfectly correlated in frequency and direction  $\bm{q}_s=-\bm{q}_i$ (assuming $\bm{q}_p=0$), although their individual  spectrum is broadband both in space and frequency. In the far field, obtained at the focal plane of a thin lens in a $f-f$ configuration, transverse mode $\bm{q}$ are focused in single transverse position $\bm{x}$ according to the transformation $(2 c f / \omega)\bm{q} \rightarrow \bm{x}$, so that momentum correlation translate in a position correlation, $\bm{x}_{s}=-\bm{x}_{i}$ (for degenerate frequency $\omega_{s}\approx\omega_{i}$). Signal and idler photons generate two symmetrical intensity noise patterns and a pairs of symmetric pixels of a camera will detect, in the same time window, the same number of photons. Quantum fluctuation affecting the object plan in the signal beam,  can be measured independently on the idler beam.
	
While the phase matching function determines the coherence time of the SPDC, which turns out to depend on the inverse of the crystal length and ranges typically in hundreds of fs, the spatial coherence in the far-field is determined by the inverse of the pump transverse size. The statistics of a single spatiotemporal mode is thermal. However, since the integration time of the detector is usually much longer than the coherence time, the measured statistics is strongly multi-thermal, with mean square fluctuation \cite{Goodman_2015}
\begin{equation}
\langle\delta^{2} n\rangle=\langle n\rangle+ \frac{\langle n\rangle^2}{M},
\label{multithermal}
\end{equation}	
where the number of detected modes $M$ is very large (here $\langle \cdot\rangle$ indicates the mean value and $\delta n=n-\langle n\rangle$). In the low gain regime, characterized by a small number of photon per mode, $\langle n\rangle/M\ll1$, even though  the number of detected photon is relatively large we can consider its statistics following a Poisson distribution with great level of approximation, with  $\langle\delta^{2} n\rangle\approx\langle n\rangle$. In this case the noise of the single beam has a white spectrum and the spatial auto-correlation is a delta function.

The detection efficiency does not affect the measured photon statistics of each beam, however it has an effect on the correlation of photon number fluctuations that is: 
	  
\begin{equation}
\langle\delta n_{s}\delta n_{i}\rangle=\eta_{c}\eta_{0}\langle n\rangle+ \eta_{c}\frac{\langle n\rangle^2}{M}\approx\eta_{c}\eta_{0}\langle n\rangle,
\label{covariance}
\end{equation}
where we have considered, for the sake of simplicity, the same detection efficiency in both channels $\eta_{s}=\eta_{i}=\eta_0$, and consequently $\langle n_{s}\rangle=\langle n_{i}\rangle=\langle n\rangle$. The efficiency contribution $\eta_{c}$ refers to the conditional probability of detecting, namely the idler photon, given that its twin signal has been detected (with the reverse being true as well). It depends strongly on the size and the correct alignment of the pair of pixels devoted to catch correlated photons.

In fact, out of the oversimplified plane wave pump assumption, the pump size determines also the width of the transverse cross-correlation. For a Gaussian distributed pump with angular full-width-half-maximum (FWHM) of $\Delta q$ the cross-correlation has still a Gaussian form $\langle \delta n(\bm{x}_{s}) \delta n(\bm{x}_{i})\rangle=\mathcal{N}\, (2\pi\sigma^{2})^{-1/2}e^{(\bm{x}_{i}+\bm{x}_{s})^{2}/2\sigma^{2}} $, where $\mathcal{N}$ is a constant that will be determinate later, with FWHM of  $\Delta x=2\sqrt{2 \log 2} \sigma=(2 c f / \omega_p)\Delta q $: if a signal photon is detected in the position $\bm{x}_{s}$ the twin idler photon will be detected according to that Gaussian probability around $\bm{x}_{i}=-\bm{x}_{s}$. It is obvious that, in order to collect most of the correlated photons, two symmetrically placed detectors (or pixels) must have areas larger than the cross-coherence area. Actually the conditional efficiency $\eta_{c}$  depends on the pixel size $L$ and on the imperfect alignment of the two pixels with respect to the optimal positions. Specifically, if the number of photons collected by the pixels in the signal and idler are $n_{s}=\int_{L\times L} n(\bm{x}_{s}) \,d\bm{x}_{s}$ and $n_{i}=\int_{L\times L} n(\bm{x}_{i}) \,d\bm{x}_{i}$, respectively, the covariance between them is: 
\begin{eqnarray}
\langle \delta n_{s} \delta n_{i}\rangle &=&\int_{L\times L} d\bm{x}_{s}\int_{L\times L} d\bm{x}_{i} \langle \delta n(\bm{x}_{s}) \delta n(\bm{x}_{i}+\bm{\Delta} )\rangle \nonumber \\
&=&\mathcal{N}\int_{L\times L} d\bm{x}_{s}\int_{L\times L} d\bm{x}_{i}\frac{1}{\sqrt{2\pi}\sigma}e^{\frac{(\bm{x}_{i}+\bm{x}_{s}+\bm{\Delta})^{2}}{2\sigma^{2}}}\nonumber \\
\label{covariance2}
\end{eqnarray}
 where $\bm{\Delta}$ is a shift vector that takes into account a possible misalignment of the two detection detectors. Comparing right hand sides of Eq. \ref{covariance} and Eq. (\ref{covariance2}), and imposing that the covariance reaches it maximum (corresponding to $\eta_{c}=1$) for  detection areas much larger than the coherence area, i.e. $\langle \delta n_{s} \delta n_{i}\rangle_{L\gg \Delta x}\approx\eta_{0} \langle  n \rangle $,  we can determine the constant as $\mathcal{N}=\eta_{0} \langle n \rangle /L^{2}$, which is nothing else than the measured photon number per unitary area. Thus the collection efficiency can be calculated as
\begin{equation}
 \eta_{c}(L,\bm{\Delta})=L^{-2}\int_{L\times L} d\bm{x}_{s}\int_{L\times L} d\bm{x}_{i}\frac{1}{\sqrt{2\pi}\sigma}e^{\frac{(\bm{x}_{i}+\bm{x}_{s}+\bm{\Delta})^{2}}{2\sigma^{2}}} 
\label{etac2}
\end{equation} 

\begin{figure}[th]
	\includegraphics[width=0.48\textwidth]{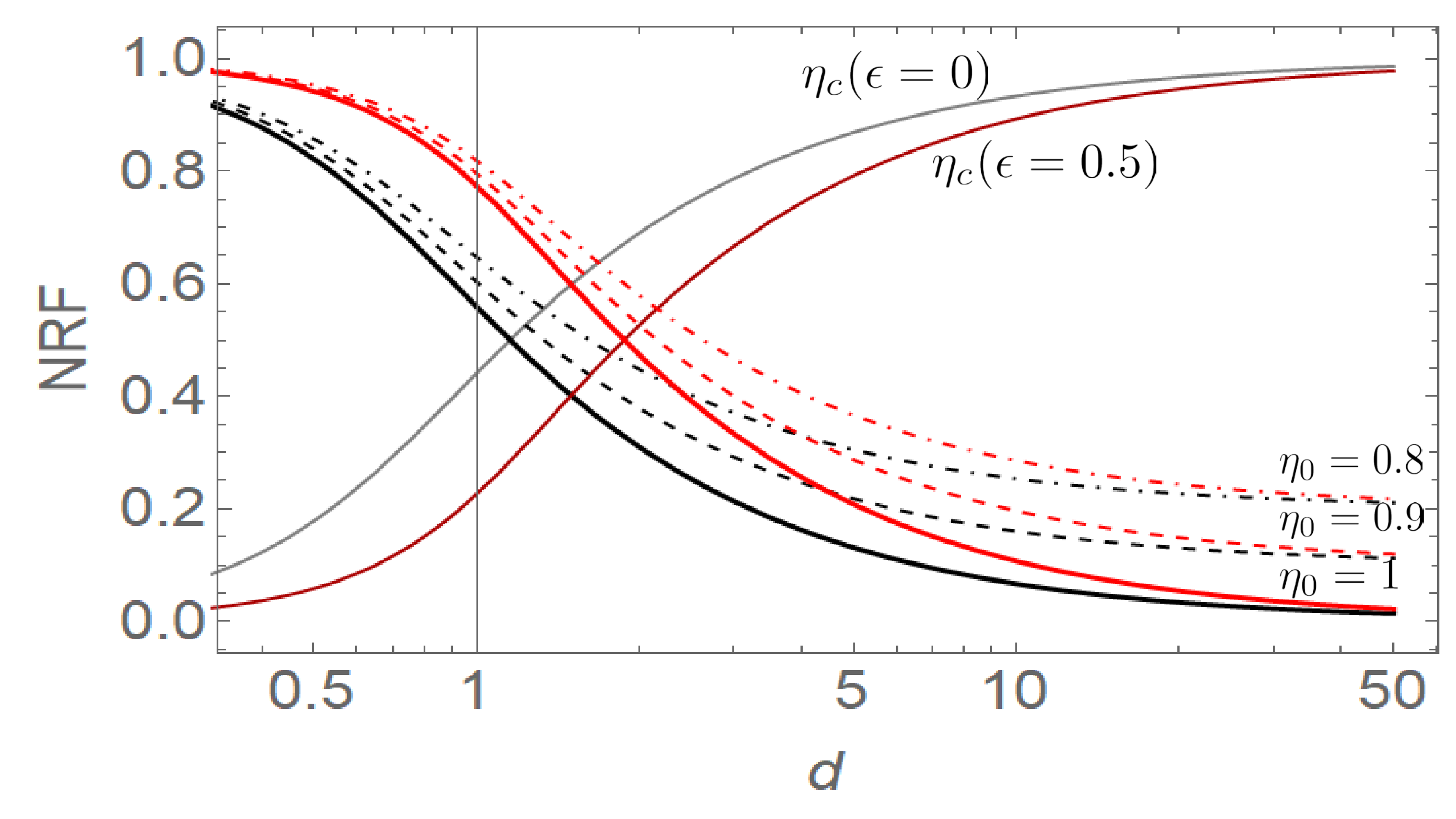} 
	\caption{Conditional efficiency and noise reduction factor as a function of the scale parameter $d$: The Gray (dark red) colored line is the conditioned efficiency in Eq. (\ref{etac2}) for a misalignment $\epsilon=0\, (\epsilon=0.5)$. The corresponding NRF, according to Eq. (\ref{NRF}), are reported in black (red) color for different values of the single channel efficiency. In particular, full line correspond to $\eta_{0}=1$, dashed line is for  $\eta_{0}=0.9$, and dot-dashed line stand for  $\eta_{0}=0.8$ }\label{NRF(d)}
\end{figure}

By a change of variable in the right hand side of Eq. (\ref{etac2}), one can easy find that the conditional efficiency turns out to depend only on the pixel size and the misalignment shift, expressed in cross-coherence lenght units, i.e. $\eta_{C}(d,\epsilon)$, where $d=L/\delta x$, and $\epsilon= \Delta/\delta x$ (we posed the same misalignment in the two directions to be the scalar $\Delta$). Fig. \ref{NRF(d)} shows  $\eta_{c}$ as a function of the scale parameter $d$. As expected for $d\ll1$ the conditional efficiency is close to zero, while for $d\gg1$ it approaches  the unit. 

Quantum correlation in the number of photons can be verified and evaluated by the so called noise reduction factor (NRF) defined as $NRF =\langle\delta^{2}(n_{s}-n_{i})\rangle/\langle n_{s}+n_{i}\rangle$ With some easy passage which exploits the single beam statistics in Eq. (\ref{multithermal}) and the covariance in Eq. (\ref{covariance}), one gets:
\begin{equation}
NRF= 1-\eta_{0}\eta_{c}+\frac{\langle n\rangle}{M}(1-\eta_{c})\approx1-\eta_{0}\eta_{c}
\label{NRF}
\end{equation}
Classical correlated state are lower bounded by $NRF\geq1$, only quantum state can reach the regime of $0\leq NRF<1$. 
As it appears from Eq. (\ref{NRF}), the NRF calculated for SPDC state is non-classical and drops to zero in case of unit efficiency. It represents a measure of the residual fluctuation compared to the shot noise level achievable by subtraction of signal and idler detected photon numbers. 
Introducing the dependence of  $\eta_{c}$ from the size and misalignment parameters $d$ and $\epsilon$ in  the expression of the NRF of Eq. (\ref{NRF}),  one realize that the effectiveness of the noise subtraction depends strongly on the spatial scale $d$, other than the single channel efficiency $\eta_{0}$. This is shown in Fig. \ref{NRF(d)}. Only at larger scale the noise can be completely suppressed, even considering a perfect alignment of the pixels. In other words, the frequency spectrum of the residual fluctuations after the subtraction is not flat. The quantum noise at lower spatial frequencies is suppressed better than the noise at higher frequencies. This is  promising, especially for phase retrieval that, as we have point out in the discussion of Eq. (\ref{noise artifact 2}), is more affected by low frequency component of the noise.

%	It is possible to measure the size of this coherence area by performing the spatial cross-correlation between the two beams:
%	$$
%	c(\xi) =  \sum_{\textbf{x}} \frac{\langle \delta \hat{N}_1(\textbf{x}) \delta  \hat{N}_2(-\textbf{x} + \xi) \rangle}{\sqrt{ \langle  [\delta  \hat{N}_1(\textbf{x})]^2  \rangle\langle[\delta  \hat{N}_2(-\textbf{x} + \xi)]^2  \rangle}}
%	$$
%	where $\xi = (x, y)$ is the shift.

\subsection*{Model for the noise reduction} 

Hereinafter, according the scheme in Fig. \ref{scheme}, the signal beam of SPDC is used as probe (P), while the idler beam is used reference (R). Thus, we will change the notation in the following way: $s\longrightarrow P$ and $s\longrightarrow R$.

Let us call $n_{P}(\bm{x})$, the unperturbed intensity pattern of the probe channel that would be detected in a certain transverse plane in absence of the object. Let also assume that, quantum fluctuation apart, the intensity is flat, meaning that the quantum mean value does not depends on  $\bm{x}$.  When object is inserted the photons are deflected from the original propagating modes creating local depletion or accumulation of photons and the pattern can be written as 
\begin{equation}\label{n'P}
n'_{P}(\bm{x})= n_{P}(\bm{x})-\delta n_{-}(\bm{x}) +\delta n_{+}(\bm{x}),
\end{equation}
where $\delta n_{-}$  represents the photons that are lost from the original path pointing $\bm{x}$ while $\delta n_{+}$ those one coming to $\bm{x}$ from the deflected modes originally pointing elsewhere.
Since the total number of photons is conserved, the spatial average of the number of photons per pixel is unchanged, i.e. $\langle n'_{P} \rangle= \langle n_{P} \rangle$ and thus $\langle \delta n_{-} \rangle= \langle \delta n_{+} \rangle$ on average. The loss of photons can be described as the action of a beam splitter of transmittance $\tau$ (average value) so that $\langle \delta n_{-} \rangle= \langle \delta n_{+} \rangle=(1-\tau)\langle n_{P} \rangle$. Here, $\tau$ represent the average of the perturbation strength induced by the phase object, with $\tau=1$ representing the absence of the object. In this work we are interested in small perturbation, that can be hidden or strongly affected by the quantum noise, so we will assume $\tau$ close to the unit.

Following the study of the optimal estimation of an absorption profile reported in our work \cite{Losero_2018,Berchera_2020}, we propose to evaluate the intensity perturbation on the probe channel as
\begin{equation}\label{I(x)}
I(\bm{x})= n'_{P}(\bm{x})- k_{opt} \delta  n_{R}(\bm{x}).
\end{equation}
The second term in Eq. (\ref{I(x)}) is meant to cancel the quantum fluctuation of the probe pattern exploiting the local correlation between probe and reference beams, and the factor $K_{opt}$ is chosen to minimize the residual fluctuation $\langle \delta^2 I(\bm{x})\rangle$, by imposing $\frac{\partial}{\partial z} \langle \delta^2 I(\bm{x})\rangle=0$. We obtain
\begin{eqnarray}\label{k_opt}
K_{opt}&=&\frac{\langle \delta n'_{P} \delta n_{R}\rangle}{\langle \delta^2 n_{R}\rangle}, \\
\langle \delta^2 I(\bm{x})\rangle&=& \langle \delta^2 n'_{P} \rangle-\frac{\langle \delta n'_{P} \delta n_{R}\rangle^2}{\langle \delta^2 n_{R}\rangle}.
\end{eqnarray} 

According to the Poisson approximation used throughout the paper, the fluctuation of $n'_{P}$ are equal to the mean value i.e. $\langle \delta^2 n'_{P} \rangle=\langle  n_{P} \rangle=\langle  n_{R} \rangle$. It represents the classical benchmark at the shot noise limit. For the calculation of the covariance in Eq. (\ref{k_opt}), note that $n'_{P}$ and $n_{R}$ are correlated only for the fraction of photons that are not deviated from the path, i.e. $\langle\delta n'_{P} \delta n_{R}\rangle= \tau \langle\delta n_{P} \delta n_{R}\rangle$, the last reported in Eq. (\ref{covariance}).

In this way, substituting in Eq. (\ref{k_opt}), we arrive to
\begin{eqnarray}\label{k_opt2}
k_{opt}&=&\tau \,\eta_{c}\,\eta_{0}, \\
\langle \delta^2 I(\bm{x})\rangle&=& \left[1-\left( \tau\,\eta_{c}\,\eta_{0} \right)^2\right]\langle n_{P}\rangle \label{noisered}
\end{eqnarray}

\begin{figure}[h]
	\includegraphics[width=0.48\textwidth]{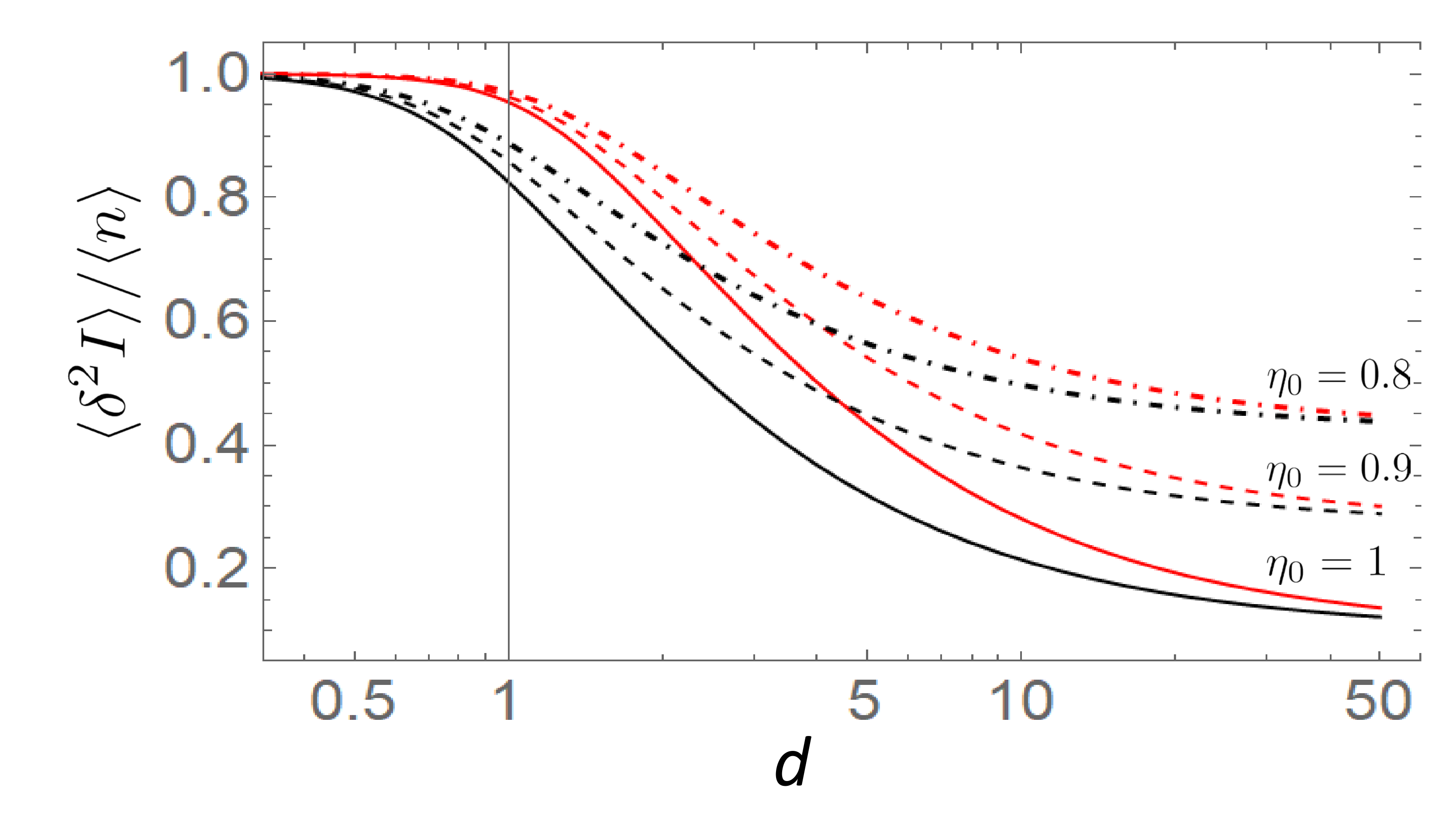} 
	\caption{Residual intensity noise in function of the scale parameter $d$:  Noise reduction of the measured intensity pattern after the correction proposed in Eq. (\ref{I(x)}) are showed in black (red) colored lines for a misalignment of $\epsilon=0\, (\epsilon=0.5)$. Full line correspond to single beam detection efficiency of $\eta_{0}=1$, dashed line is for  $\eta_{0}=0.9$, and dot-dashed line stand for $\eta_{0}=0.8$ }\label{dI2}
\end{figure}

As expected the correlation between probe and reference beam can reduce the shot noise in the perturbed intensity noise pattern with respect to the case of the measurement on the probe channel only. As usual, the noise reduction depends on the detection efficiency, not only of the single channel but also from the conditional efficiency $\eta_{c}$  of detecting pairs of correlated photons in corresponding pairs of pixels. As discussed after Eq. (\ref{covariance}) the dependence of the conditional efficiency from the elementary resolution area makes the frequency spectrum of the noise not flat, allowing a better suppression of the noise at lower frequencies (higher spatial scale). The dependence from the scale factor and from the single beam detection efficiency is shown in Fig.  \ref{dI2}.  
In the following we will describe simulation where the TIE in Eq. \ref{tie} is solved using the detected intensity pattern after the noise correction as in Eq. \ref{I(x)}. 

\section{Results}	
	\begin{figure*}
	\centering
		\includegraphics[width=1\textwidth]{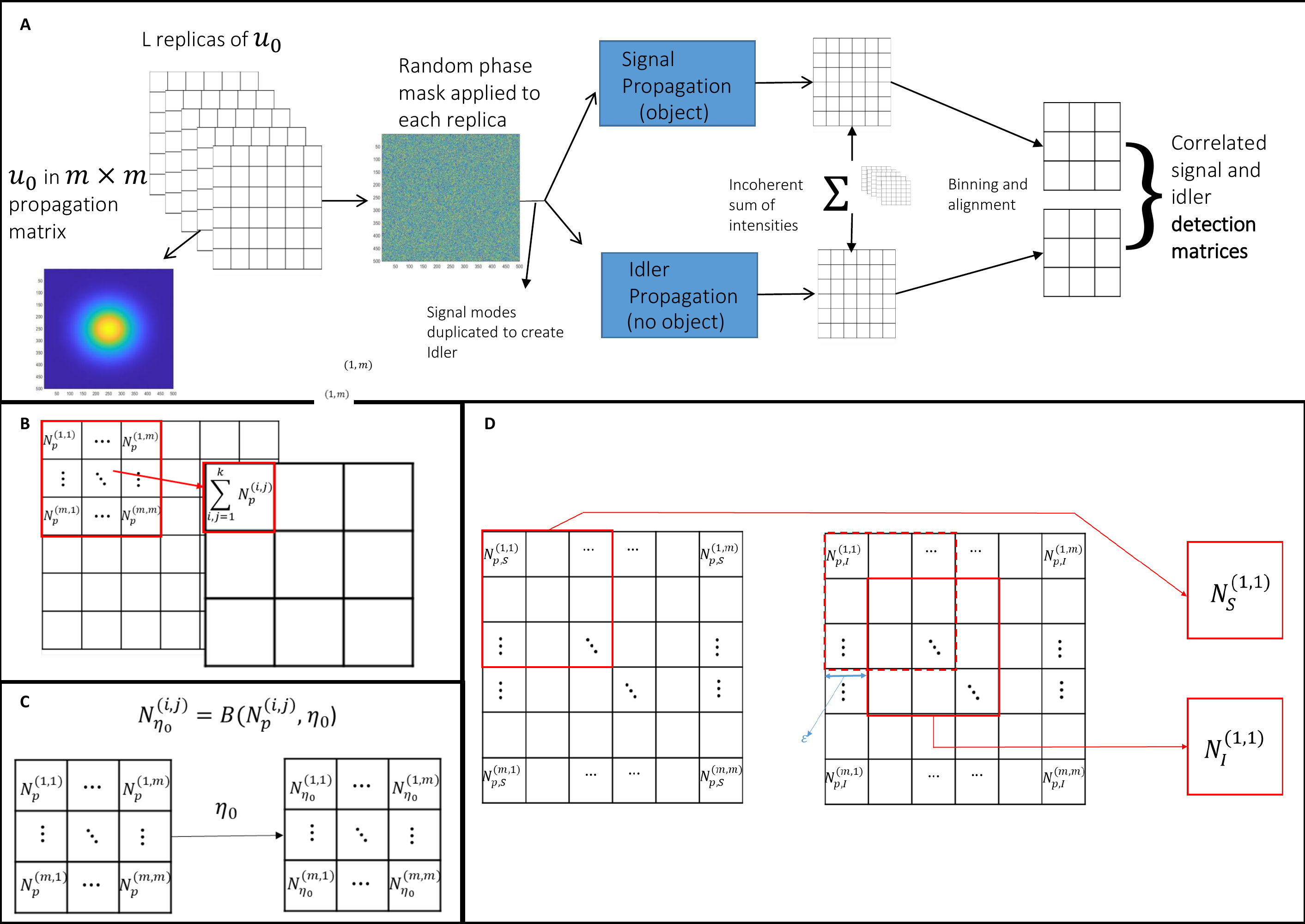} 
		\caption{\textbf{A.} \emph{Simulation scheme.} $L$ replicas of an initial field $u_0$ are created each with a random phase profile to simulate incoherence. Each of the modes is propagated independently through the QCPR setup of Fig. \ref{scheme} (in the configuration of interest). The final intensity profiles are obtained by an incoherent sum of the intensities of the $L$ modes at the detection plan. Finally correlated noise is added to signal and idler profile and a binning is performed on the propagation matrix to yield the detection matrix. \textbf{B.} \emph{Binning.} The detection matrix is obtained by a binning operation where the values of the pixels in each $k_b\times k_b$ square are added to obtain the value of the binned pixel. \textbf{C.} \emph{Efficiency simulation.} The detection efficiency $\eta_0$ is simulated by performing for each pixel with value $N$ an extraction from a binomial distribution with $N$ trials and probability of success $\eta_0$. \textbf{D.} \emph{Conditioned efficiency.} The conditional efficiency $\eta_c$ is simulated by introducing a shift of $\epsilon$ between the binning of signal and idler profiles. }\label{exp}
	\end{figure*}	

\subsection*{Phase retrieval Simulation }
The QCPR scheme in Fig. \ref{scheme} relies on two main physical properties of the probing light source: partial coherence on the probe beam and quantum correlations between probe and reference field intensities. The simulation of this scheme can be done with a "semi-classical" approach. First we perform a "classical" step consisting in the generation and propagation of a partially coherent beam to the imaging plane, and, at a second time, we introduce the shot noise. Correlation between probe and reference beams are simulated by producing two deterministic identical copies of the beams, up to the object plane. Also the shot noise at the image (detection) plane is reproduced identical in the probe and reference. Only the effect of the simulated non unit detection efficiency, produces a deterioration of the correlation among probe and reference intensity noise. After the intensity pattern has been determined on focus and the two defocussed plans at $\pm \delta z$, the phase profile can be recovered using the TIE equation, that can solved numerically in different ways \cite{Weller_2014,Zuo_2020}
\subsubsection*{Signal Propagation}
To simulate the partial coherence, we use at the source plane a collection of $L$ independent field modes of size $m\times m$ pixels, with identical Gaussian intensity profile and randomized phase (delta function correlated). This is depicted in Fig. \ref{exp}.\textbf{B}. Each incoherent masks is propagated along the optical system, composed by the far field lens, the phase object in the proximity of the far field (at distances $z=0,\pm\delta z$) and the equivalent of a $2f-2f$ imaging lens. The final intensity pattern in the image plane is obtained by the incoherent sum of all the modes contributions. This methodology to simulate partial coherence propagation in described in detail in \cite{Voeltz_2011}. 
According to the Van Cittert-Zernike theorem \cite{VanCittert_1934,Zernike_1938}, the coherence function, $\gamma(\textbf{x},\textbf{x'})$, at the far field is the Fourier transform of the source intensity, so that at the far field plane, we obtain a field, $u(\textbf{x})$ with an almost flat intensity profile, $ I(\textbf{x})=\langle u^*(\textbf{x})u(\textbf{x})\rangle$ and Gaussian coherence function  
\begin{equation}
\gamma(\textbf{x},\textbf{x'})= \frac{\langle u^*(\textbf{x})\;u(\textbf{x}') \rangle}{\sqrt{I(\textbf{x})\;I(\textbf{x}')}}=e^{-\frac{r^2}{2 l_c^2}} \label{coh},
\end{equation}
where $r=|\textbf{x}-\textbf{x}'|$ and the coherence length $l_c$ is related to the width $w$ of the  Gaussian field at the source plane by the equation 
\begin{equation}
	l_c= \frac{f\lambda}{\sqrt{2}\; w \pi}. 
\end{equation}

where $f$ is the focal length of the far field lens and $\lambda$ the wavelength. The intensity profile will be flat on average with  multithermal statistics \cite{Goodman_2015}. In fact, each single simulated field mode represents a single temporal mode of the real source. In the far field it produces a classical spatial excess noise at the coherence length scale, known as speckle pattern, where the variance of intensity fluctuation are proportional to its square. The independent contributions of the $L$ masks reduces the excess noise of a factor $L$.

\subsubsection*{Detection matrix and correlated idler system}
In the correlated scheme of Fig. \ref{scheme}, the the cross-correlation length between the signal and the idler is the same as the auto-correlation of the single beam intensity. The noise reduction needs, as showed in the previous section, spatial bandwidth of the detector larger than the inverse of the correlation length. In other words, pixel size should by larger enough to collect mostly correlated photons. Thus, the need arises to have a distinction between the \emph{propagation} matrix and the \emph{detection} one.
On one hand, an accurate simulation requires that the propagation 'pixel' is much smaller than the relevant spatial features, namely the one set by a coherence area. 
%This is because, in this scenario, the sampling error due to the discretization would be no longer negligible and the effects of the partial coherent propagation would risk to be misrepresented. 
On the other hand, in the physical system the noise subtraction is efficient only on spatial scales larger than the coherence area, as it has been discussed after Eq. (\ref{noisered}). To set the spatial scale at the detection, one can use different approaches. The simplest one consists in the application of a binning of size $d \, k_b$ to the propagated  $m \times m$ intensity matrix at the image plane, where $k_{b}$ is the number of propagation pixels corresponding to a coherence length at the image plane, and $d$ is the scaling factor introduced in after Eq. (\ref{etac2}). For example, if $d=1$, one binned pixel correspond to one coherence area. In our simulation we have approximatively $k_{b}=5$. The binning process is represented in Fig. \ref{exp}.\textbf{B}. 
%The pixels of the propagation matrix are grouped in binned pixels each containing $k_b \times k_b$ propagation pixels. The value of each binned pixel is the sum of the intensity of each of the propagation pixels contained in it.  The end result is the detection matrix having dimension $m_b \times m_b$, $m_b=m/k_b$.

For the statistical analysis that follows it is convenient to use the number of photons, $N$, in place of the intensity, $I$, and since both detection time and area are fixed, the two quantities are proportional. The only noise source rising in the classical propagation are the already mentioned multi-thermal statistical fluctuations, which scale as $\langle N \rangle^2 $. In many quantum imaging experiments using a PDC source, the number of temporal modes collected at detection is very large and the mean number of photons per mode is very small ($\langle N \rangle/L<<1$). In this conditions the multithermal noise $\sim\langle N \rangle^2/L$ is negligible with respect to the shot noise contribution $\sim\langle N \rangle$. To mimic this experimental conditions we set a number of random mask much larger than the average number of detected photons (see Fig. \ref{exp}). However, because of the long computational time each propagation requires, we were limited to consider $L\sim10^{3}$ temporal modes, thus the number of photon per detection pixel should be kept relatively small too, $\langle N \rangle\leq10^{2}$. 

Signal and idler beam are generated using the same set of random phase masks, so that their classical multi-thermal pattern are identical, with the only difference that the reference (idler) beam does not interact with the phase object during the way, and does not contain phase induced perturbation. To include in our simulation the effect of quantum shot noise, we add poissonian noise manually to the $m\times m$ image matrix of the signal and a perfectly correlated one to the idler system. The single channel efficiency, $\eta_0$, introduced in the previous section, is simulated by extracting the number of photons in each pixel from a binomial distribution: for each pixel at position $(i,j)$, having $N^{(i,j)}_p$ photons we  extract the new number of photons $N^{(i,j)}_{\eta_0}$, from the binomial distribution $B(N^{(i,j)}_p,\eta_0)$ (see also Fig. \ref{exp}.\textbf{C}). This operation is carried independently for signal and idler pattern so that, as a result, some of the correlation between signal and idler intensity patterns is lost. 

The last element of the physical system that we need to simulate is the noise reduction depending on the spatial frequency, as reported in Fig \ref{dI2}. As discussed in the "Materials and Methods" section, one factor that limits the conditional efficiency $\eta_c$, is the misalignment $\epsilon$ (in unit of coherence area) of the pixel pairs devoted to detect correlated photons in the signal and idler detection plane. It turns out that, the introduction of a proper shift of the reference matrix of a quantity $\Delta_{pix}=\epsilon \,k_{b}$ generates approximatively the requested scale dependent conditional efficiency and thus the requested properties of the noise reduction, when the probe noise pattern is corrected by the shifted reference one. This step is showed in Fig. \ref{exp}.\textbf{D}).

\subsection*{Simulation Results} 
	\begin{figure*}
		\centering
		\includegraphics[width=\textwidth]{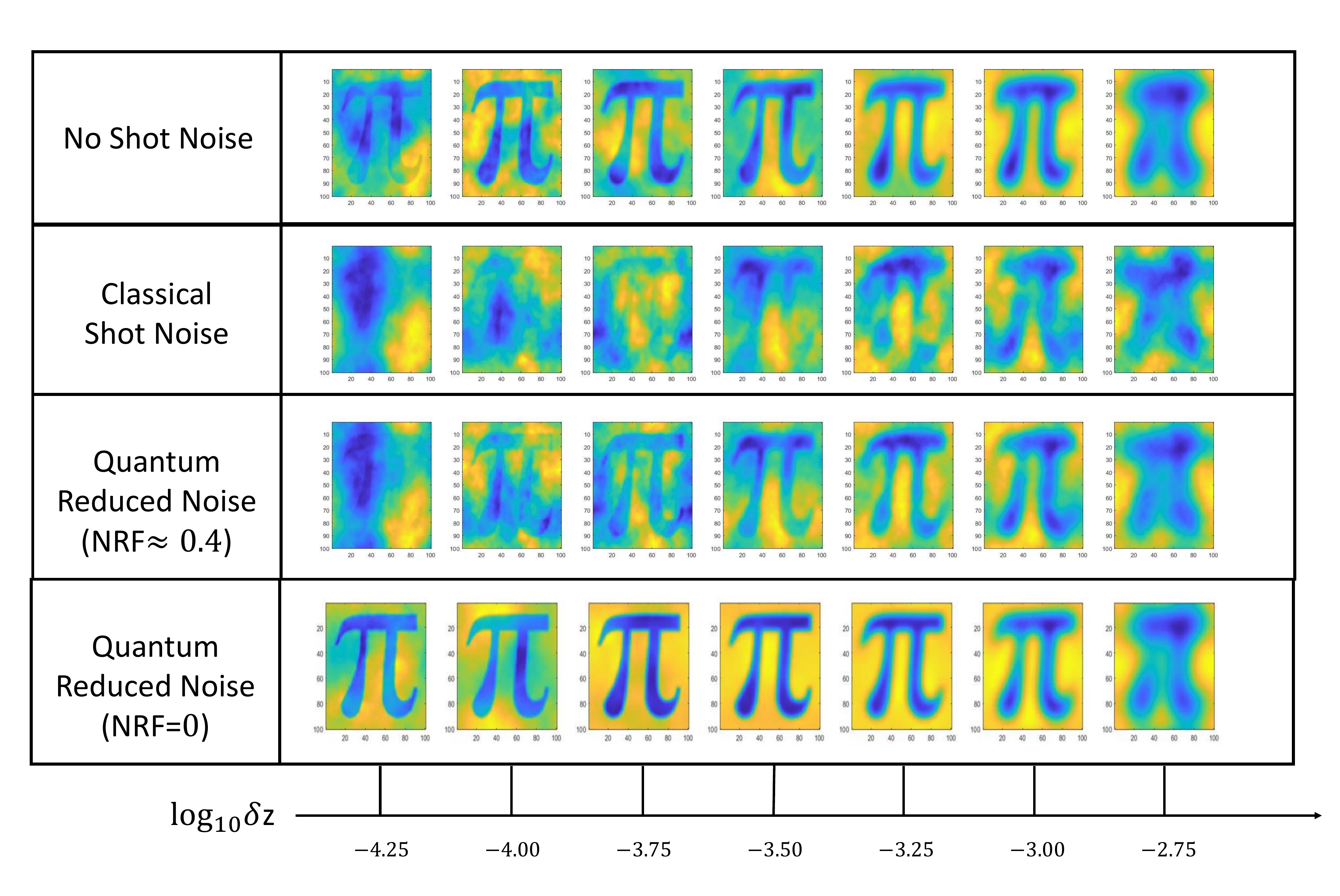} 
		\caption{\emph{Noise reduction simulation}. Reconstructions of the phase mask reported in the inset of Fig. \ref{res2} (height of the phase step equal to $\pi/4$) are pictured for different $\delta z$ and different noise of the probe state. The $\delta z$ are reported in meters and are in log scale. The probe used in the simulation has a coherence length of $l_c=7.2\cdot 10^{-6}$ m, the focal length of the far field lens is $f=10^{-2}$ m and the size of the matrix at the object plane is set to $6\cdot10^{-4}$ m. In the first row shot noise is not added to the probe so that the only source of noise are small multithermal fluctuations. In the second one poissonian shot noise is added. In the third row the shot noise is added and subsequently reduced using correlations with a NRF$\approx0.4$. In the fourth row the noise reduction on the shot noise is performed assuming perfect correlations (NRF$=0$).}\label{res1}
	\end{figure*}
		\begin{figure}
		\centering
		\includegraphics[width=0.45\textwidth]{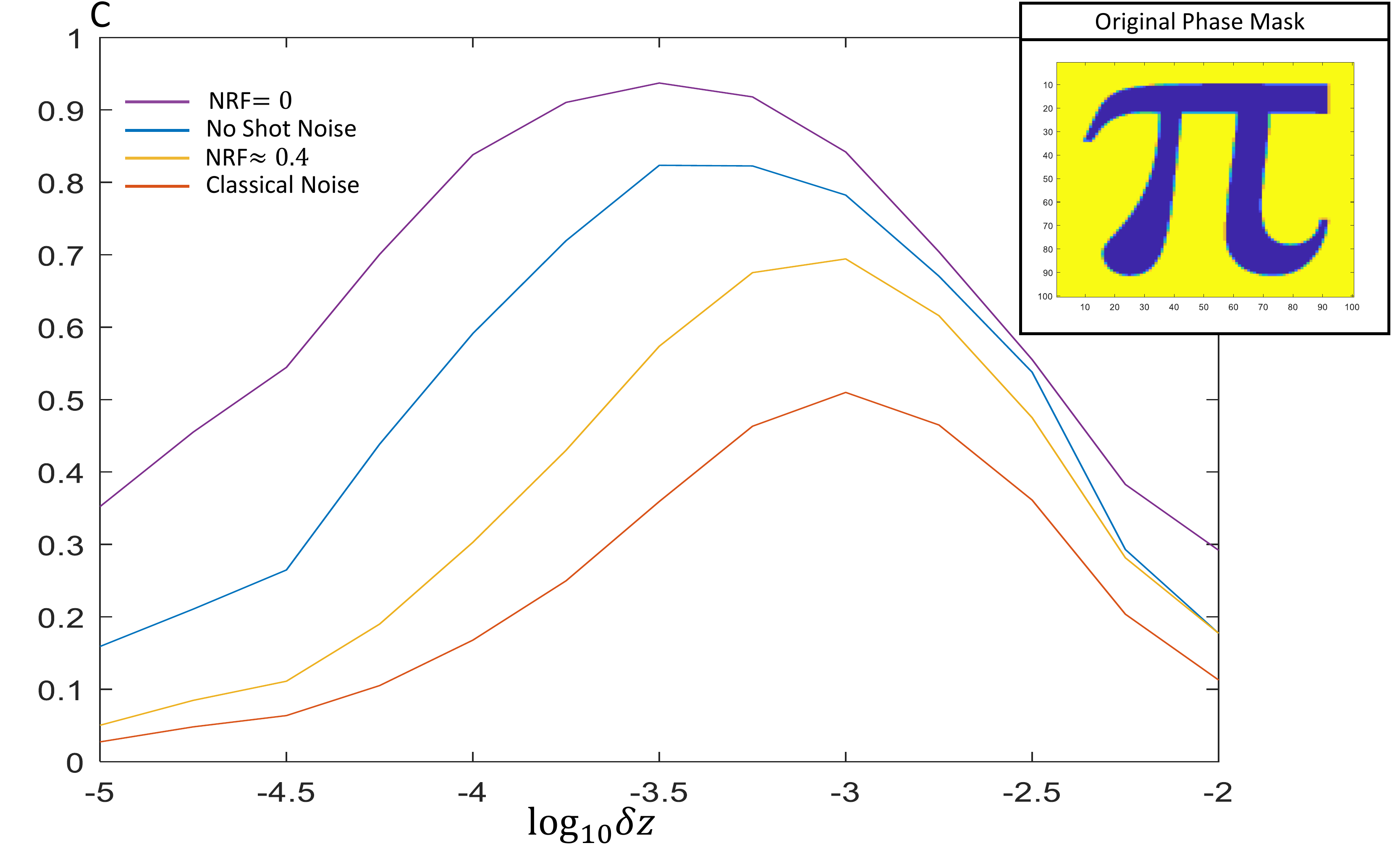} 
		\caption{\emph{Reconstruction quality as a function of $\delta z$}. In the plot is showed the correlation coefficient, of the reconstructed phase with the original one (reported in figure), quantifying the quality of the reconstruction, as a function of the defocus distance $\delta z$. The difference in phase between the letter $\pi$ and the background is $h=\pi/4$. The blue line refers to the situation in which no shot noise is added to the probe (see first row of Fig. \ref{res1}) . The red one to a probe with shot noise (see second row of Fig. \ref{res1}). Finally the yellow and the purple ones refer to the case in which the shot noise has been reduced by quantum correlations, respectively with NRF $\approx0.4$ (yellow, see third row of Fig. \ref{res1}) or NRF $=0$ (purple, see fourth row of Fig. \ref{res1}). }\label{res2}

	\end{figure}
		\begin{figure}
		\centering
		\includegraphics[width=0.45\textwidth]{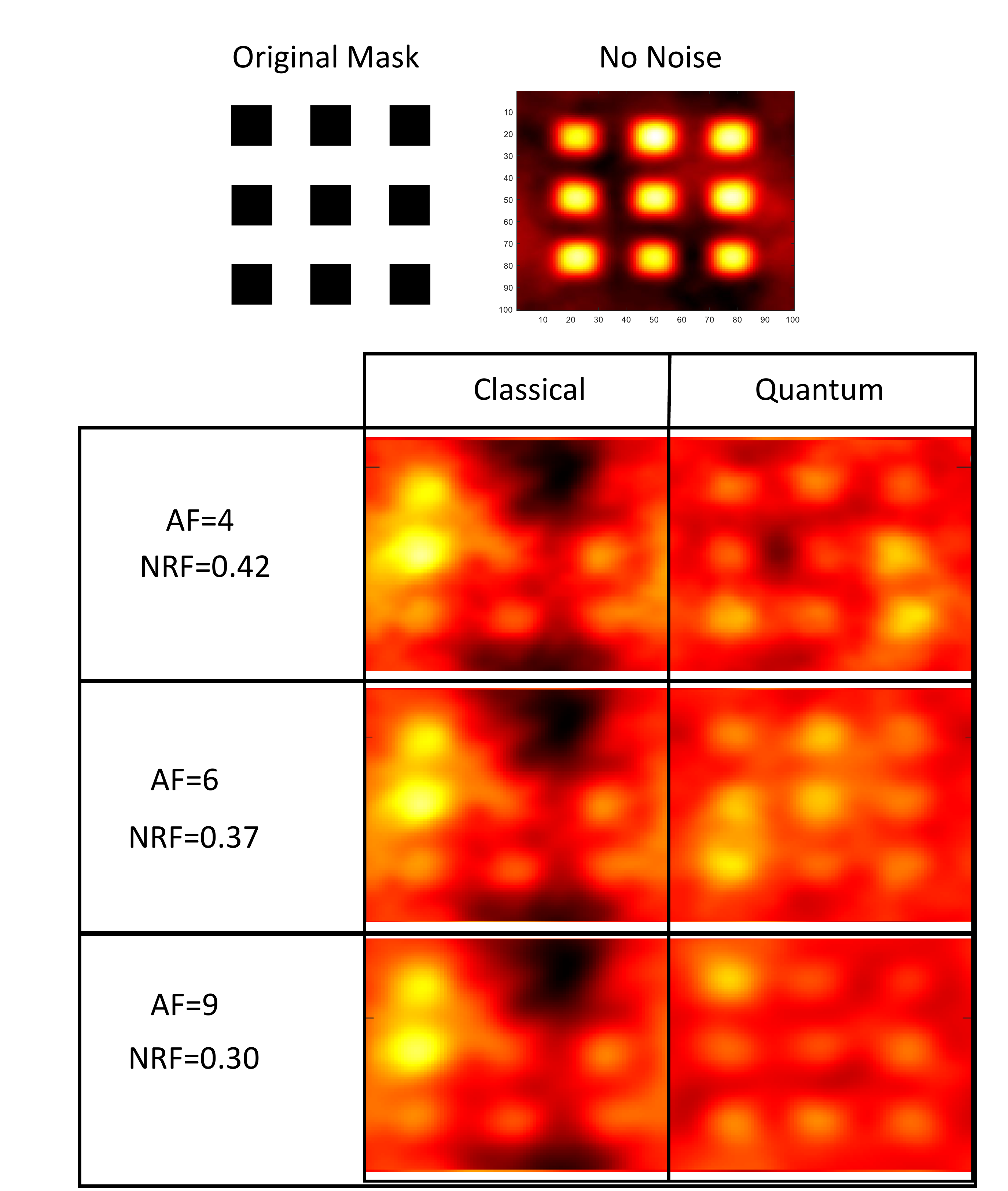} 
		\caption{\emph{Averaging Filter (AF) reconstructions}. In figure are shown reconstructions of a grid of nine squares (whose phase difference with the background is $h=\pi/8$), considering a classical probe (shot noise) and a quantum one (reduced noise). In the first row the reconstructions are obtained applying to the intensity matrices a MF of size $k=4$ resulting in a NRF=0.42. In the second and third the size of the AF applied are respectively $k=6$(NRF=0.37) and $k=9$(NRF=0.3).}\label{res3}
	\end{figure}
The results of the simulation described in the previous section are reported in Fig. \ref{res1}, where the reconstruction of the phase is shown as a function of the defocussing distances $\delta z$.  
The first line refers to the phase retrieval obtained only using the intensity measurement performed on the probe beam, in absence of shot noise contribution. 
Even in this case the reconstruction is noisy due to well known \textit{deterministic} effects, namely the sampling error due to discretization of the image, affecting especially the reconstruction at small $\delta z$ and the finite difference approximation of the derivative in $z$, influencing the retrieval at large defocussing.  This defines a range in $z$, even without random noise, for which the reconstruction is optimal.
Moreover, the multi-thermal excess noise, although very small (less than 5\% of the signal) gives some contribution to the overall quality of the reconstruction. Note that, in the experimental condition of many realizations with SPDC, this contribution would be negligible, even if it could be substituted by similar effects coming from detector electronic noise.

The quality of the reconstruction can be evaluated quantitatively in terms of the \textit{correlation coefficient}, between the original phase profile $\phi$ and the reconstructed one $\phi_r$:
\begin{equation}
\mathcal{C}= \frac{\sum_{i,j}(\phi_{r}(i,j)-\bar{\phi_{r}})(\phi(i,j)-\bar{\phi})}{\sqrt{\text{Var}[\phi_{r}]\text{Var}[\phi]}}
\end{equation}
where $\bar{\phi}$ and Var$[\phi]$ denote the spatial mean and variance of the matrix $\phi$. For the noiseless case the dependence of $C$ from $z$ is showed in Fig. \ref{res2} (blue line) where it can be seen that the correlation is in agreement with the eye test. 

In the second row of Fig. \ref{res1} are showed the reconstruction when poissonian shot noise is introduced at the detection. It yields a drop in the quality of the reconstruction for all values of $\delta z$. In particular the region of smaller $\delta z$ is the most affected, since the intensity variation due to phase gradients in this region are smaller and, as a result, the signal to noise ratio (SNR) is lower. We note however how the effect of random noise is still very evident in the $\delta z$ range that should be optimal without random noise. The result is that the best reconstructions in this situation are obtained for bigger $\delta z$ than in the noiseless case, in a region where the derivative approximation noise starts to be noticeable, especially losing higher spatial frequency. This effect can also be seen in Fig. \ref{res2} where the correlation coefficient $\mathcal{C}$ for reconstruction done with shot noise is reported in red. 

The third row of Fig. \ref{res1} reports the reconstructions when the shot noise has been reduced using quantum correlations between probe and reference, according to Eq. ($\ref{noisered}$). In particular, we set the parameters in the simulation as $\eta_{0}=0.95$ and $\epsilon= 0.25$ (corresponding to $\eta_c=0.64$) that gives a noise reduction factor of $NRF \approx 0.4$ at the resolution showed of $100 \times100$ pixels. There is a clear improvement for all reconstruction due to the noise reduction. Moreover, smaller $\delta z$ are now accessible due to increased SNR leading to an improvement in the reconstruction of higher spatial frequency. This advantage is also showed in Fig. \ref{res2}. 
Finally, the last row of Fig. \ref{res1}, is obtained when the noise cancellation is perfect, i.e. when both excess classical noise and shot noise are considered perfectly correlated. Note that, it would correspond to a unit single beam detection and conditioned efficiency. This is not achievable experimentally but we use this limit case to show that, even in the total absence of noise, the sampling resolution and the finite difference approximation of the derivative in $z$ does not allows a accurate reconstruction at all defocussing values. The comparison with this ideal case allows also a correct estimate the role of the shot noise. Quantitatively, one should compare the red and purple curves in Fig. \ref{res2}. 

We observe that, the $NRF \approx 0.4$ used to produce the third row of Fig. \ref{res1}, while obtained in wide field SSNQI experiment at lower resolution (lower $m$), in practice may be hard to get at the resolution shown in Fig. \ref{res1}. While obtaining a $\eta_0$ close to $0.95$ is in principle feasible with actual technology, it may be difficult to obtain $\eta_c=0.64$ for a $100 \times 100$ detection matrix due to the requirements on the size parameter $d$. An useful operation that can be done to effectively increase the scaling parameter $d$ while maintaining the same dimension of the detection matrix, instead of binning, is to apply the \emph{averaging filter}. Using $k$ sized averaging filter, to each pixel is assigned the value of the average of its $k-$neighbor pixels. If the original size parameter of the image is $d$ after the averaging filter operation its \emph{effective} size parameter will be  $d'=k d$. The dimension of the reconstruction matrix will be preserved, although part of the information encoded in the higher frequency is lost. The effect of the average filter, with realistic parameters for an experimental implementation (NRF $\approx 0.8$ at resolution $100\times 100$), is reported in Fig. \ref{res3}. On the first row the reconstruction, with classical shot noise and quantum noise reduction, are performed after the application of a median filter of size $k=4$, yielding a NRF=$0.42$. The classical image is heavily affected by the noise and very little information on the initial image is recovered. On the other hand the quantum enhanced one, simulated with parameters feasible  in an experimental realization, recovers the pattern with an increase in precision and removing the heavy noise-induced artifacts characterizing the classical reconstruction. The effect is even more evident in the second and third row where the size of the median filter is increased to $k=6$ (NRF=$0.37$) and $k=9$ (NRF=$0.30$), where the quantum recovery is improved while there is no evident effect on the classical side. 

Fig. \ref{res3} shows how the use of adequately sized averaging filter is useful in noise dominated scenarios --this is the case for example of faint phase profile with low energy probing-- to improve the reconstruction. This is done by reducing the NRF at the cost losing finest details. It also shows how a real quantum advantage can be obtained with current technology, suggesting an experimental implementation in the near future.

\section{Conclusions}
In this paper we propose a new scheme, that we dubbed as QCPR, to enhance a non-interferometric recovery of a phase profile using quantum correlations in photon numbers. In our proposal the TIE is used to recover the phase while an ancillary system is used to perform noise reduction. The result is a sensible improvement in the quality of the phase recovery. Our results also show how an advantage can be obtained in experimental conditions, considering realistic parameters modeling various experimental inefficiencies limited by current technology. We also show that the use of a averaging filter results in an increase in the effectiveness of the noise reduction. An experimental realization of our protocol is underway. Our proposed scheme can find application in different fields. A particular suitable application would be to biological imaging, where many sample are quasi-transparent and encode a good amount of information in the phase. Moreover in those situations its not unusual to have limitations on the probing energy, as high energy could damage the samples, so that, due to the low SNR, noise reduction would be even more important.

\section{Acknowledgments}
We acknowledge the supporting Project 17FUN01 ’BeCOMe’ within the programme EMPIR. The EMPIR initiative is co-founded by the European Union’s Horizon 2020 research and innovation programme and the EMPIR Participating Countries. 
		
\bibliography{bib}		
\end{document}